\documentclass[aps,reprint, nofootinbib, showpacs]{revtex4-1}
\usepackage{graphicx}
\usepackage{graphicx, epsfig}
\usepackage{color,soul}
\usepackage{mathrsfs}
\usepackage{amsmath}
\usepackage{amssymb}
\usepackage{hyperref}

\newcommand{\beq}{\begin{equation}}
\newcommand{\eeq}{\end{equation}}
\newcommand{\bea}{\begin{eqnarray}}
\newcommand{\eea}{\end{eqnarray}}

\newcommand{\ben}{\begin{enumerate}}
\newcommand{\een}{\end{enumerate}}

\newcommand{\be}{\begin{equation}}
\def\bel#1{\begin{equation} \label{#1}}
\newcommand{\ee}{\end{equation}}
\newcommand{\bi}{\begin{itemize}}
\newcommand{\ei}{\end{itemize}}
\newcommand{\ba}{\begin{align}}
\newcommand{\ea}{\end{align}}
\newcommand{\comments}[1]{}
\usepackage{color}
\definecolor{cblue}{RGB}{100,5,255}
\definecolor{cred}{RGB}{255,50,40} 
\definecolor{cgreen}{RGB}{40,255,40} 
\definecolor{corange}{RGB}{250,200,40} 
     
\begin{document}
\title{Attractor Models in Scalar-Tensor Theories of Inflation}
\author{Sukannya Bhattacharya}
\email[]{sukannya.bhattacharya@saha.ac.in}
\affiliation{Theory Divison, Saha Institute of Nuclear Physics, HBNI,1/AF Bidhannagar, Kolkata - 700064, India }
\author{Kumar Das}
\email[]{kumar.das@saha.ac.in}
\affiliation{Theory Divison, Saha Institute of Nuclear Physics, HBNI,1/AF Bidhannagar, Kolkata - 700064, India }
\author{Koushik Dutta}
\email[]{koushik.dutta@saha.ac.in}
\affiliation{Theory Divison, Saha Institute of Nuclear Physics, HBNI,1/AF Bidhannagar, Kolkata - 700064, India }


\begin{abstract}

In this work we study the cosmological attractor models of inflation in connection with certain scalar-tensor theories of gravity, e.g $f(R)$ gravity and Brans-Dicke theory. For some particular choices of the functional degrees of freedom in these theories, one obtains Starobinsky like predictions in the ($n_s$-$r$) observable plane. We have demonstrated that these choices in the Lagrangian density of certain $f(R)$ and Brans-Dicke theories fulfil the condition of the cosmological attractors. That explains why known predictions of $f(R)$ and Brans-Dicke theories in certain cases appear to be the predictions of the much discussed attractor theories. In addition, we did an analysis showing how the predictions of an attractor model is preserved with respect to the variation in the functional freedom of the theory.

\end{abstract}
\maketitle

%
\section{Introduction}
Inflationary paradigm has emerged as the leading candidate for the explanation of cosmological structures that we see today \cite{inflation1, inflation2}. The fluctuations in the cosmic microwave background (CMB) as observed by the Planck satellite can be explained by a nearly scale-invariant primordial spectrum parametrised by its amplitude and scalar spectral index $n_s$ \cite{planck2015}. The very nature of inflation also produces primordial metric fluctuations whose amplitude is parametrised with respect to the scalar amplitude, and it is denoted by tensor-to-scalar ratio $r$.

According to the Planck 2015 data the constraints on these observable parameters are $n_s = 0.968 \pm 0.006$ with $r < 0.11$ \cite{planck2015}. Interestingly, the observable predictions of Starobinsky model $R + R^2$ \cite{inflation1}, the model with a non-minimal coupling $\xi \phi^2 R$ and $V(\phi) \sim \phi^4$ \cite{Bezrukov:2007ep, non_minimal}
fall into the sweet spot of the Planck 2-$\sigma$ contour. For a large number of e-folds $N$, the observables for these models are given by 
\begin{equation} \label{attarctor_observation}
n_s = 1 - 2/N, ~~~~ r = 12/N^2 
\end{equation}
where $N = 50$-$60$ is the time when the CMB scales leave the horizon during inflation\footnote{Unconventional post inflationary dynamics can affect the preferred number of e-folds, and thus inflationary observables. For example, see \cite{moduli}.}. In terms of a canonically normalised scalar field in minimal gravity, all the above mentioned models have exponentially flat potential in large field values. 

Among large varieties of potentials, the models with a plateau like behaviour are generically favoured by the recent data \cite{Ijjas:2013vea}. Subsequently, interests renewed in understanding models that can produce small gravitational wave with spectral index in the above mentioned limit. Using Lyth bound, that relates the value of $r$ with the field excursion $\Delta \phi$ during inflation, these type of models would require $\Delta \phi \lesssim \mathcal{O}(M_{Pl})$ \cite{Lyth:1996im}. It has been achieved in two ways. Firstly it was noticed that a coupling  between the inflaton with heavy fields can effectively flatten the inflaton potential \cite{Dong:2010in}. Secondly, Kallosh and Linde discovered a class of models whose observable predictions are attracted towards the point of Eq.~\eqref{attarctor_observation} when one parameter in the model is continuously modified \cite{Galante:2014ifa}. In fact, the models with arbitrarily small $r$ were also proposed. The predictions of Starobinsky model just sit at this attractor point. It was found that these class of attractor models have underlying conformal symmetry structure, and their supergravity realisations have been discussed extensively in the literature \cite{sugra_conformal_origin}. 

One type of attractor model, namely the $\alpha$-attractor, finds its attractor nature from the second order pole in the kinetic energy term \cite{Kallosh:2013yoa}. In terms of the field variable, the potential must be smooth at the position of the pole. 
 In this case, the potential of the canonically normalised field asymptotes to a constant value. In canonical field, the pole is shifted to infinity and so is never reached physically through its dynamical evolution during inflation. We get a nearly shift-symmetric plateau in the asymptotic limit. For the case of $\xi$-attractor models, the gravity is non-minimally coupled to the scalar field. In this case with the proper choice of the non-minimal coupling function, the kinetic term, and the potential function, the predictions of the model quickly converge to the asymptotic value given by Eq.~\eqref{attarctor_observation} as we increase $\xi$. Both these attractor models can be unified in the picture of kinetic formulation of the theory, where $n_s$ is related to the order of the pole of the kinetic term in its Laurent series expansion and $r$ primarilly depends both on the leading order pole and also on the residue corresponding to that pole in the expansion. 

In this work, we have studied $f(R)$ theories of gravity and Brans-Dicke theory in the context of attractor models for inflation. We show explicitly how these models can be rewritten in terms of the attractor models with appropriate kinetic term that is suitable for attractor mechanism to work. For some particular choice of the functional degrees of freedom in these theories, one obtains Starobinsky like predictions in the $n_s$-$r$ observable plane. Any choice of these functions fixes the potential in the Jordan frame or in the Einstein frame in terms of the non-canonical field. Whether any model would show attractor properties crucially depends on the asymptotic nature of these potential functions. For example, any deviation from $R + R^2$ gravity distorts the asymptotic nature of the potential, and makes the potential unstable for attractor. Our work is complementary to the approach taken by Ref \cite{Broy:2014xwa} where the effects of the asymptotic shift symmetry breaking corrections to the potential corresponding to $R^2$ term have been studied. Similar studies have been also carried out in \cite{Huang:2013hsb,fR_relation,LindePlanck13}.

We also discuss the robustness of attractor mechanism by varying conformal function in the case of $\xi$-attractor, and analyzing the effect of higher order pole in the kinetic energy term. In the case of $\xi$ attractor, even when the conformal function is changed by adding higher order monomial, for a sufficiently large value of $\xi$ the predictions come back to the attractor point. On the other hand, when the kinetic function is changed with a higher order pole, the predictions deviate from the attractor curve. This is consistent with the conclusion of \cite{Broy:2015qna} where changes to the observables have been calculated in the limit of perturbative corrections to the kinetic function.

We emphasise the point that even when a model has a kinetic term with a suitable pole structure (as we will recast), the potential in the Jordan frame is fixed from the underlying structure of the model. Only for certain functional choices we get the attractor behaviour. In the case of Brans-Dicke theory, we choose only these functions judiciously, and show how predictions for the attractor models are guaranteed to be reproduced when certain limits of the model parameters are taken. 

The motivation of this paper is two-fold: \\
$ \bullet $ After reviewing the explicit mechanism of the attractor dynamics, we check the robustness of this mechanism for higher order corrections in the functional degrees of freedom for $\alpha$ and $\xi$- attractors.\\
$ \bullet $ We theorize how attractor mechanism can be obtained from scalar-tensor theories and explicitly show the $f(R)$ theory and Brans-Dicke theory as examples. For this, we study the allowed range of the parameters in the functional degrees of freedom in light of the observables in CMB.\\
We will see that the plateau nature of the effective potential for large fields values in attractors is governed by the dynamics of the functional forms of the non-minimal couping (for $\xi$-attractors) and the pole-containing kinetic term (for $\alpha$-attractors). Therefore, both the analyses mentioned above relate to this dynamics and its outcomes.

This paper is organized as follows. In the next section we summarize the attractor mechanism. In section \ref{Sensitivity_of_attractor_models}, we analyze the robustness of the attractor mechanism. In section \ref{fr_attr_section} , we discuss $f(R)$ inflation models in the language of attractor models, and discuss phenomenology when polynomial terms are present in the action. In section V, we discuss Brans-Dicke theory in the context of attractor models, and find potentials which suitably provide attractor solutions. Finally, we conclude in section VI. 

\section{Attractor mechanism for inflation models}
\label{attractor-disc}

A class of inflationary models has been identified whose predictions in the space of observables are not so sensitive to the specific potential function due to particular non-canonical nature of the kinetic term. The data coming from PLANCK experiment shows this coincidence amongst various inflation models like - Starobinsky model \cite{inflation2}, Goncharov-Linde Model \cite{Goncharov:1983mw}, supersymmetric version of non-minimal chaotic inflation with $\phi^4$ potential \cite{Ferrara:2010yw,Linde:2011nh,Kallosh:2013wya} and Higgs inflation \cite{Bezrukov:2007ep}. In the leading approximation of $1/N$, where $N$ being the number of e-folds of inflation, the observable predictions \emph{i.e.} scalar spectral index $(n_s)$ and tensor-to-scalar ratio $(r)$ of all these models are attracted to a common point given by Eq.~\eqref{attarctor_observation}. These models are collectively known as cosmological attractors. 

The cosmological attractors broadly come into two categories, where the Lagrangian either has a non-minimal coupling to the Ricci scalar or may feature a characteristic kinetic term with a second order pole. The former description is known as non-minimal $\xi$-attractor and the later one is called $\alpha$-attractor, where $\xi$ or $\alpha$ is a free dimensionless parameter of the theory which when varied the predictions converge to Eq.~\eqref{attarctor_observation}. The origin of the attractor properties of both kinds can be traced back to the pole structure of the Kinetic function in its Laurent expansion and the potential function is smooth at the position of that pole \cite{Galante:2014ifa}. Under special condition these models can be mapped to each other.

\subsection{$\alpha$ and $\xi$ attractors}
The Lagragian for the models of cosmological attractor is given by 
\begin{equation} \label{attarctor_kinetic}
\mathcal{L} = \sqrt{-g_E}\left [ \frac{1}{2} R_{E} - \frac{1}{2} \left(\frac{a_p}{\phi^p} + \dotsm \right)(\partial \phi)^2 - V_E(\phi) \right]~,
\end{equation}
where the kinetic function is given by a Laurent series expansion with a pole of order $p$ at $\phi = 0$ (without loss of generality), and the dots denote subleading terms. We approximate the potential energy by a Taylor series expansion $V_E(\phi) = V_0(1 + c \phi + \dotsm )$ near the vicinity of the pole. Here $\phi$ is the inflaton field with non-canonical kinetic energy term, and gravity is minimally coupled. The constant $V_0$ sets the asymptotic value of the potential in term of the canonical field.

It turns out that the observable predictions of this model are uniquely characterised by the properties of the pole. In particular, the scalar spectral index $n_s$ and the tensor-to-scalar ratio $r$ at leading order in $1/N$ are given by \cite{Galante:2014ifa} 
\begin{equation}
\label{predcitions_general}
n_s = 1 - \left(\frac{p}{p-1} \right) \frac{1}{N}, ~~~~~ r = \frac{8 c^{\frac{p-2}{p-1}} a_p^{1/(p-1)}}{(p-1)^{\frac{p}{p-1}}}\frac{1}{N^{\frac{p}{p-1}}}~.
\end{equation}
Note that whereas the spectral index depends only on the order of the pole, the tensor-to-scalar ratio depends both on the order and the residue of the pole. For $p = 2$ and $a_p = 1$, this yields the famous Starobinsky inflation prediction for the scalar spectral index given by Eq.~\eqref{attarctor_observation}. Depending on the value of $a_p$, the tensor-to-scalar ratio can be arbitrarily small. The second order pole with $p = 2$ is special as its origin can be traced back to some superconformal supergravity theories \cite{sugra_conformal_origin}, and to non-minimal gravity theories in the Jordan frame \cite{Kallosh:2013tua}.  

As mentioned earlier, there are primarily two classes of cosmological attractors. Both of them can be interpreted as having a pole in the kinetic term of order $p =2$. The Lagrangian for the cosmological $\alpha$-attractor is given by,
\begin{align}
\mathcal{L}=\sqrt{-g_E}\bigg[\frac 1 2 R_E-\frac1 2\frac{\alpha(\partial\phi)^2}{(1-\phi^2/6)^2} -\alpha f^{2}(\phi/\sqrt{6})\bigg]~,
\label{alpha_attractor}
\end{align}
where $\alpha$ is a dimensionless parameter of the model. 
If we make a field redefinition as $\phi/\sqrt{6} = (1-\rho)/(1+\rho)$, we can write the above Lagrangian as\footnote{Instead of making this field redefinition, if we make the Laurent series expansion of the kinetic function $K_E$, at the leading order we get a second order pole at $\phi = \sqrt{6}$ with a residue of $3\alpha/2$. The subleading terms in the expansion do not contribute to the observables in the large $N$ limit.}
\begin{align}
\mathcal{L}=\sqrt{-g_E}\bigg[\frac{1}{2} R_E - \frac{3\alpha}{2\rho^2}\frac{(\partial \rho)^2}{2} - \alpha f^{2}(\rho) \bigg]~.
\end{align}
This is  similar to what is written in Eq.~\eqref{attarctor_kinetic} with the specified form of the kinetic function with $a_2 = \frac{3}{2}\alpha$. Therefore, the Lagrangian of an $\alpha$-attractor model also features a second order pole at $\rho =0$.  

In terms of the canonical field $\hat \phi$, the pole at $\rho = 0$ is shifted to large field values, and the potential is given by 
\begin{align}
V_E(\hat \phi) = \alpha f^{2}\left[\tanh(\hat \phi/\sqrt{6\alpha})\right]
\end{align}
where $\rho = e^{-\sqrt{\frac{2}{3\alpha}} \hat \phi}$. For monomial functions, the potential reduces to the form of T-models of conformal attractors. For the choice of $f(x) = \frac{cx}{1+x}$, one finds the generalization of Starobinsky potential \cite{Galante:2014ifa} 
\begin{align}
V_E = \frac{\alpha c^2}{4} \left(1 - e^{-\sqrt{\frac{2}{3\alpha}}} \hat \phi \right)^2~.
\label{starobinsky_pot}
\end{align}
This potential has a long plateau at large $\hat \phi$. It is this particular functional form of $V_E$ that makes the potential asymptotically flat at large values of the canonically normalized field. The predictions of this model varies from  quadratic chaotic inflation model (for large $\alpha$) to Starobinsky model (for $\alpha = 1$) with \cite{Kallosh:2013yoa} 
\begin{equation} \label{attarctor_observation_general}
n_s = 1 - 2/N, ~~~~ r = 12\alpha/N^2 ~.
\end{equation}
This model can also produce arbitrarily small $r$ for $\alpha \ll 1$.

Note that the choice of the function $f(x)$ can be more general than what has been mentioned above. Because of the nature of $\tanh(\hat \phi)$ function, when the argument of the function becomes of order one, the potential is stretched with a constant asymptotic plateau. But the function must be chosen appropriately such that the post inflationary vacua is consistent with observations. 

The other description for the cosmological attractor with non-minimal coupling to gravity is given by \cite{Kallosh:2013tua, non_minimal_1}\footnote{For multifield models of inflation with non-minimal couplings, the expressions for $n_s$ and $r$ in the leading approximation in $1/N$ is different~\cite{Kaiser_multi}. Higher order correlation functions for these models are studied in~\cite{sayantan_cosmos-e}.} 
\begin{align}
\mathcal{L}_J=\sqrt{-g_J}\bigg[\frac 1 2 \Omega^2(\phi)R_J - \frac 1 2 K_J(\phi)(\partial\phi)^2 - V_J(\phi) \bigg] ~,
\label{attr_jr}
\end{align}
where $\Omega^2(\phi)=1+\xi f(\phi)$ is the conformal factor. Here the theory is defined in a Jordan frame. The corresponding Einstein frame description after a conformal transformation of the metric tensor $g_{\mu\nu}^E=\Omega^2(x^\mu) g_{\mu\nu}$ is 
\begin{align}
\mathcal{L}_E = \sqrt{-g_E} \bigg[\frac{1}{2}R_E - \frac1 2\bigg(\frac{K_J}{\Omega^2}+ 6 \frac{\Omega'^2}{\Omega^2}\bigg)(\partial\phi)^2 -  V_{E} \bigg] ~.
\label{ein_att}
\end{align}
Here $V_E = V_J/\Omega^4$, and the prime is w.r.t the field variable $\phi$. If $K_J(\phi) \ll 6\Omega'^2$, the above Lagrangian reduces to the usual form of the attractor model.
\begin{align}
\mathcal{L}_E = \sqrt{-g_E} \bigg[\frac{1}{2}R_E - 3 \frac{(\partial \Omega)^2 }{\Omega^2} - V_E(\Omega) \bigg].
\label{ein_att_approx}
\end{align}
In this case, the canonically normalised field $\hat \phi$ is related to the conformal factor by $\Omega^2 = e^{\sqrt{\frac{2}{3}} \hat \phi}$.
Now, with $\xi$ being negative, $\Omega^2(\phi)=1+\xi f(\phi)$ has a pole of order two in the kinetic term. However, with positive $\xi$, the pole structure is clear if we make the transformation $\Omega^2 \rightarrow 1/\rho$, and the above Lagrangian becomes
\begin{align}
\mathcal{L}_E = \sqrt{-g_E} \bigg[\frac{1}{2}R_E - \frac3 4 \frac{(\partial \rho)^2 }{\rho^2} - V_E(\rho) \bigg].
\label{ein_att_approx_rho}
\end{align}
Thus, at large $\xi$ or large $\phi$, the pole at $\rho \rightarrow 0$ is accessible. If $V_E(\rho)$ is smooth at the position of the pole, the potential w.r.t to the canonically normalised field is going to be flattened for large field values. Therefore both the classes of cosmological attractors are basically a realisation of the same description given in Eq.~\eqref{attarctor_kinetic} through redefined field variables. For the case of $\xi$-attractor, the residue at pole is $a_2 = 3/2$, and the predictions are given by Eq.~\eqref{attarctor_observation} for a suitable choice of $V_J$.
For the particular choice of $V_J(\phi)= c^2 (\Omega^2-1)^2$ this yields the famous Starobinsky model. On the other hand, for the choice of $V_J=\Omega^4f^2\big(\frac{\Omega^2-1}{\Omega^2+1}\big)$, the resulting theory in the Einstein frame becomes a T-model of $\alpha$-attractor \cite{T-model}. See \cite{Racioppi_linear} for some other attractor models where the form of the potential function has to be approximately close to this form to show attractor mechanism.

In summary, any choice of the potential function of the non-canonical field having second order pole in the Einstein frame will show attractor nature if it satisfies the following three criteria:\\
$ \bullet $ Potential must be a smooth function at the location of the pole,\\
$ \bullet $ the potential has to be a positive definite function, and \\
$ \bullet $ at large field values, potential must asymptotically approach to a constant (or nearly constant) value.\\ 
The second criteria is a statement about the boundedness of the potential from below. The last one is seemingly significant as this        asymptotic domain of the potential is responsible for the attractor type predictions in the $n_s$-$r$ plane. 

We note that the way the Einstein frame scalar potential of a canonical field manifesting an asymptotically long plateau is slightly different for the two kinds of attractor models. In the case of $\alpha$-attractor, the potential in the Einstein frame after canonical normalization is controlled by the canonical conversion function $\tanh{\hat{\phi}}$, and it causes flattening of the potential. It happens because in the defining Lagrangian of the $\alpha$-attractor, the pole in the Kinetic term appears at some finite value of $\phi$. Nevertheless, a suitable field redefination can make the pole to appear at zero field value. But in that case the potential still remains a $\tanh{\hat{\phi}}$ function in the canonical field. Therefore, for the $\alpha$-attractor viewing the pole in non-canonical field either at zero or at finite value does not make any difference in the argument of the potential function of the canonically normalized field. In contrast, for the $\xi$-attractor, the canonical conversion generates an exponential function $\Omega^2 \sim e^{\hat{\phi}}$. But in this case, a tacit choice of the potential function $V_J$ makes the potential exponentially flat. However, in either case the asymptotic behaviour of the Einstein frame scalar potential is similar. In summary, in addition to the pole structure in the kinetic energy term, the attractor behaviour also crucially depends on the properties of the potential functions. 


In section~\ref{fr_attr_section} we are going to study how the inflationary predictions of some well studied scalar-tensor theories can be reinterpreted in the language of attractor models. 

\section{Robustness of attractor mechanism}
\label{Sensitivity_of_attractor_models}

The attractor mechanism works due to the existence of a second order pole in the kinetic term, and the potential being smooth at the position of the pole. As we have seen in the previous section, for the case of $\xi$-attractor, a certain condition has to be satisfied between the conformal factor and the potential function. Specifically, the order of the monomial in both these functions must be the same such that asymptotically the potential in the Einstein frame becomes constant. On the other hand, the shift symmetric potential in the asymptotic limit can be broken by including higher order poles. These higher order poles in general can appear when the Kinetic function is expanded in Laurent series \cite{Galante:2014ifa,Broy:2015qna}. For the non-minimal $\xi$-attrator, higher order corrections arise when conformal function is Taylor expanded \cite{Kallosh:2013tua}. In this section, we will analyze the robustness of the attractor mechanism by modifying the the conformal factor\footnote{Note that here we modify the conformal factor perturbatively which is different from quantum corrections to the attractor models, which is studied extensively in \cite{Fumagalli_qc}.} and the pole structure of the kinetic function. If the corresponding corrections arise at field values much larger than the field value $\phi_{60}$ when the CMB scales goes outside the horizon, the attractor prediction remains robust. For the case of perturbative corrections to the leading order pole in the kinetic function, the corrections to the inflationary observables have been shown to be universal \cite{Broy:2015qna}. We analyze this case when the corrections are not necessarily perturbative.


Here we would like to see how a correction term in the non-minimal coupling function is going to affect an otherwise attractor like predictions. For the purpose of our analysis we start with the Lagrangian density of Eq.~\eqref{attr_jr}. For simplicity, the non-minimal function and the potential function of this theory are taken to be
\begin{align}
\Omega^2(\phi) = 1+\xi(b_1\phi +b_2\phi^2),  \quad\qquad V_J(\phi) = m^2\phi^2,
\end{align}
where $b_1,b_2$ are arbitrary constants of the theory. Recasting the Lagrangian density into the Einstein frame through a conformal transformation we obtain for the potential function to be
\begin{align}
V_E(\phi) = \frac{m^2\phi^2}{[1+\xi(b_1\phi +b_2\phi^2)]^2} ~.
\label{vesens}
\end{align}
With $b_2=0$ and after eliminating $\phi$ in terms of $\Omega^2$, the potential can be written as $V_E = \frac{m^2}{\xi^2}(1-\Omega^{-2})^2$. This represents a $\xi$-type attractor model with predictions interpolating between quadratic chaotic model and Eq.~\eqref{attarctor_observation} when $\xi$ is varied from zero to large values. Here $\xi$ is the attractor parameter when its value is increased. Without loss of generality, for the purpose of our analysis we have taken $b_1=1$ as it can be absorbed in the redefinition of $\xi$. We now want to see how this attractor like behaviour changes when we include $b_2\phi^2$ term in the non-minimal function. 

Let us now investigate the predictions of this potential in the light of PLANCK 2015 data. With the form of the Einstein frame potential specifield, one can calculate the scalar spectral index and the tensor-to-scalar ratio from the following expressions respectively,
\begin{align*}
n_s &=1-6\epsilon_E +2\eta_E,  \qquad r=16\epsilon_E 
\end{align*} 
where $\epsilon_E$ and $\eta_E$ are the inflationary slow roll parameters which are given as,
\begin{align}
\epsilon_E &= \frac1 2 \bigg(\frac{V_E'(\hat{\phi}) }{V_E}\bigg)^2, \qquad   \eta_E = \frac{V_E''(\hat{\phi})}{V_E} 
\end{align}
These parameters have to be calculated when observable CMB modes go outside the horizon. 

\begin{figure}[h!]
    \includegraphics[height=6.8cm]{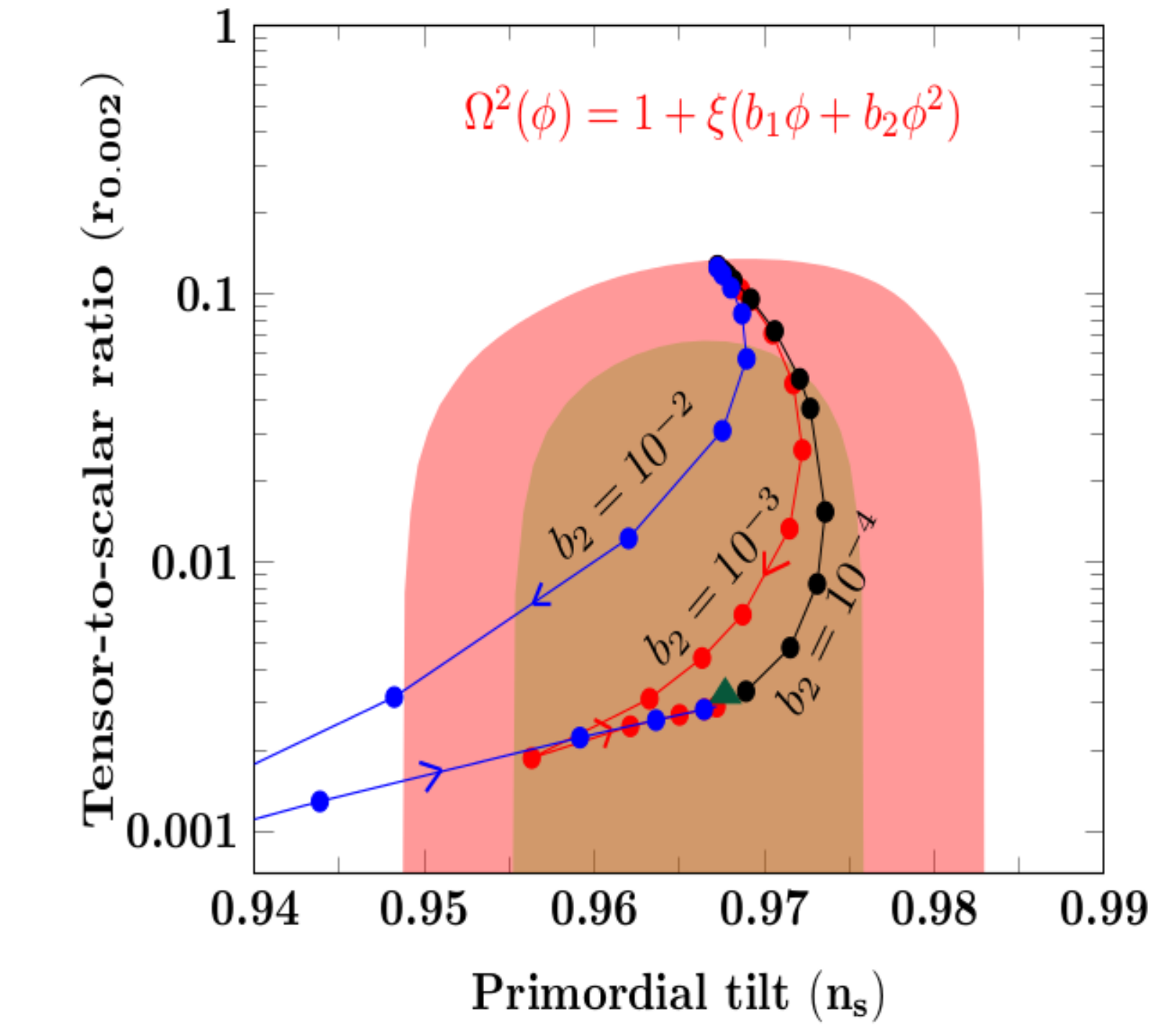} 
             \caption{Plot showing the variations of $n_s$-$r$ with the $68\%$ and $95\%$ confidence contours from 2015 Planck data. The arrow in each line shows the direction of increasing $\xi$. The green triangle in the plot shows the attractor point given by Eq.~\eqref{attarctor_observation}.} 
\label{nsr_sensti}
\end{figure}

The theory now contains three free parameters $m,\xi$ and $b_2$. But $m$ gets fixed from the amplitude of curvature perturbation, and $\xi$ and $b_2$ remain free parameters that we vary. For several representative values of $b_2$, we change $\xi$ from zero to $10^4$ and calculate the scalar spectral index and the tensor-to-scalar ratio. The observable predictions in this case are shown in Fig.~\ref{nsr_sensti}. 
In this figure the various colored curves corresponds to fixed values of the $b_2$ parameter. All of them approach to the attractor point in the $n_s$-$r$ plane from their $\xi = 0$ limit of quadratic chaotic inflation limit. The rightmost black curve with $b_2 = 10^{-4}$ goes directly into the attractor point. But the way other two curves with larger value of $b_2$ (red, blue) approaches towards the attractor point is quite different from the first. With increasing value of the $\xi$-parameter they initially deviate from the attractor point. Thereafter, for further increase of $\xi$, the curves once again return to the attractor point. To understand this behaviour let us have a close look at the expression for the Einstein frame potential in Eq.~\eqref{vesens}. There is a maximum of the potential at $\phi=\phi_0 = \frac{1}{\sqrt{b_2\,\xi}}$ when the effect of $b_2\phi^{2}$-term in the denominator is comparable to $\phi$-term. Note that this is true for any non-zero value of $b_2$ with $\phi_0 \rightarrow \infty$ when $b_2 \rightarrow 0$. The potential now develops two branches. For values of $\phi>\phi_0$, it asymptotes to zero while it acquires a flat part for $\phi<\phi_0$. For viable inflation we must have $\phi_{60} <\phi_0$, and necessary inflation can proceed in the flat part of the potential.  

Now the value of $\phi_{60}$ depends on both $\xi$ and $b_2$. For smaller values of $b_2 \sim \mathcal{O}(10^{-4})$, the value of $\phi_{60} \gtrsim 1$, but due to the smallness of $b_2$ parameter the effect of the quadratic term is negligible, and the curve directly moves towards the attractor point. On the other hand for $b_2 \sim \mathcal{O}(10^{-3})$, the curve initially moves away from the attractor point for up to a certain value of $\xi$. In this case, up to the turning point, even though $\phi_{60} \gtrsim 1$, the effects of larger $b_2$ in the quadratic correction term is appreciable. With further increasing value of $\xi$, the potential distorts and inflation happens with $\phi_{60} \lesssim 1$ in an asymptotic flat part. In this case, the quadratic term becomes negligible in the region where inflation proceeds, and the attractor point is reached. In conclusion, we find that if we increase the value of $\xi$ sufficiently, we regain the attractor behaviour even with the correction term with $b_2 \sim \mathcal{O}(1)$ in the conformal factor. 
We therefore conclude that the attractor behaviour is very robust to the perturbations in the functions that define the attractor Lagrangian as long as the attractor parameter is increased sufficiently. 

We now would like to understand how an attractor theory is sensitive with respect to the variations in its kinetic function. For this analysis, we pick up the Einstein frame Lagrangian density given in Eq.~\eqref{attarctor_kinetic}. The kinetic and the potential functions are taken as\footnote{For a general discussion on poles of higher orders in relation to attractor models, see \cite{Terada:2016nqg, Broy:2015qna}}
\begin{align}
K_E(\phi) = \frac{a}{\phi^2} + \frac{b}{\phi^3}, \qquad V(\phi) = V_0 (\phi-1)^2~.
\end{align}
From Eq.~\eqref{predcitions_general}, we know that the attractor point in $n_s$ vs. $r$ space is reached for a pole of order two with $b = 0$. We are now going to study what happens to the curve reaching to the attractor point when a third order pole is additionally present. In the absence of third order pole term, the attractor parameter is $a$ when its value is decreased. For $a = 1$, we reach Starobinsky point. 

\begin{figure}[h!]
    \centering
    \includegraphics[height=6.8cm]{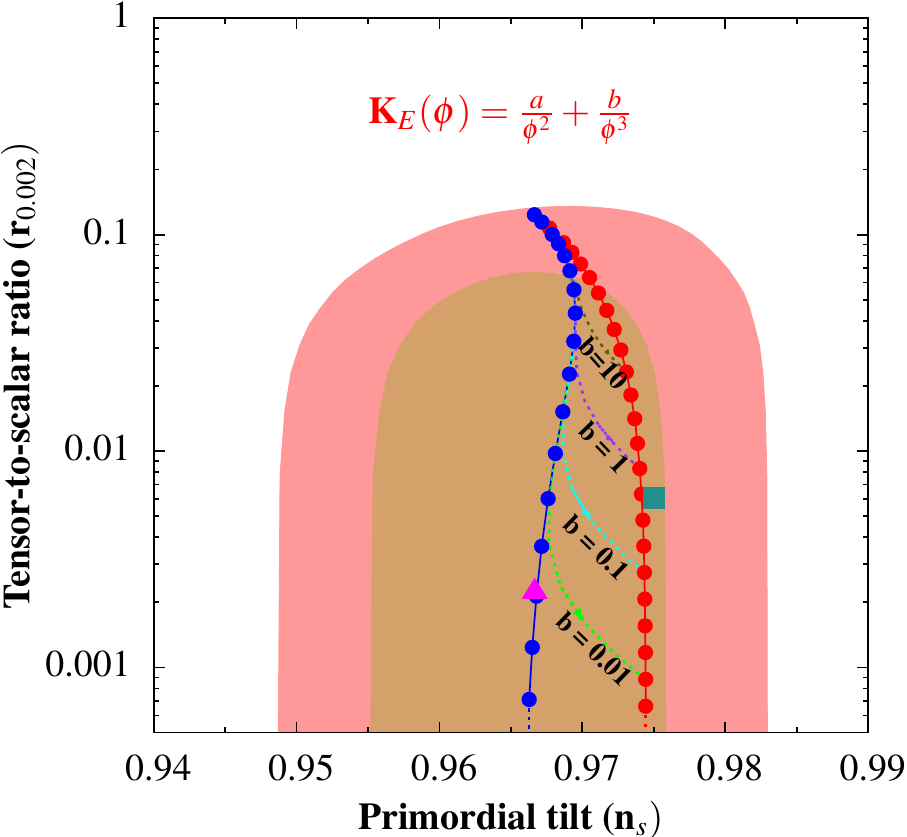} 
             \caption{Plot showing the variations of $n_s$-$r$ with the $68\%$ and $95\%$ confidence contours from 2015 Planck data. The blue dots correspond to the quadratic pole only, and the red dots correspond to the cubic pole. The arrow in each dotted lines shows the direction of decreasing $a$ for various $b$ values. The magenta triangle  and the blue square correspond to Eq.~\eqref{predcitions_general} with $p=2$ and $p=3$ respectively for $a_p = 1$.} 
\label{kinetic_sensiti}
\end{figure}

Here both $a$ and $b$ are free parameters of the theory, and $V_0$ will be fixed from the amplitude of scalar curvature perturbations. To perform the analysis we have kept the parameter $b$ fixed at some representative values. For each fixed value of $b$ we change $a$ from large values of order $\sim 10^4$ to small values and calculate the inflationary observables. The predictions of this model are  shown in Fig.~\ref{kinetic_sensiti}.
In this figure the blue curve on the left represents the prediction when $b=0$ \textit{i.e.,} there is only second order pole in the kinetic function. The curve approaches the attractor point (magenta triangle) with decreasing value of $a$ (from top to bottom), and it is consistent with general predictions of Eq.~\eqref{predcitions_general}. The red curve on the right shows the predictions for having only the third order pole $(a=0)$ in the kinetic function. Between these two curves the various (dotted) curves show how inflationary predictions are changing when poles of both orders are present in the theory. For $b=0.01$ (dotted green curve) the value of $\phi_{60}\sim\mathcal O(10^{-3})$, and in this case initially the effect of $b/\phi^3$-term is subdominant compared to $a/\phi^2$-term. But with gradual lowering of attractor parameter $a$, $\phi_{60}$ remains constant and the third order pole starts to affect the observables. Finally, with a sufficient small value of $a$, the predictions finally hits the line for only having third order pole in $K_E$. For other values of $b$, this behaviour remains the same.   


In all these dotted curves, it turns out that after a certain critical value $a \lesssim a_0$ (say), when the curves start to deviate from the blue line, the value of $\phi_{60}$ at first decreases. After that with further decreasing $a$, $\phi_{60}$ practically becomes unchanging. 
Therefore, the strength of the cubic pole becomes dominant over the quadratic pole. 
For certain non-zero values of $b$, even though some curves pass through the attractor point, they never come back to the same point with further decreasing the attractor parameter $a$. 
%
In summary, it is the dependence of $\phi_{60}$ upon the attractor parameter ($\xi$ for the case modifying the conformal factor, and $a$ for modifying the kinetic term) that determines whether the predictions will approach back to the attractor point or not.


\section{$f(R)$ theory as attractor models} 
\label{fr_attr_section} 

The modification of Einstein's theory of gravity is an interesting avenue in exploring physics beyond the standard picture of Big Bang cosmology. Because of high curvature in the early universe during inflation, the corrections to the Einstein-Hilbert gravity turns out to be generic \cite{inflation1}. In general these corrections are such that either the geometry can be non-minimally coupled to some scalar field or higher derivative term in the metric can appear. Study of these higher derivative theories are important when gravity is quantized in a curved spacetime background and the issue of renormalization is addressed \cite{Birrell:1982ix,Gottlober:1989ww}. Moreover, they also appear in studies of infation in early universe \cite{Huang:2013hsb}.In its simplest version, the corrections may take the form of some arbitrary function of the Ricci scalar $R$. The action for this modified theory of gravity is given by \cite{DeFelice:2010aj},
\begin{align}
S=\frac{1}{2\kappa^2} \int d^4x \sqrt{-g}f(R),
\label{fR_lagr}
\end{align} 
where $\kappa^2 =8\pi G=1/M_{pl}^2$ and the Ricci scalar $R=g^{\mu\nu}R_{\mu\nu}$ is the contracted version of Ricci Tensor $R_{\mu\nu}$. 
Each choice of the function $f(R)$ corresponds to a different theory and a large number of viable theories exist in the literature for both late time and early universe cosmology. Out of the diverse possibilities of $f(R)$, from the standpoint of inflationary cosmology the form $f(R) = R+R^2$, proposed by Starobinsky, grew with alluring attention for its remarkable agreement with Planck observations.

By a conformal transformation $g_{\mu\nu}^E=\Omega^2(x^\mu) g_{\mu\nu}$ of the metric tensor, the above theory can be recasted in the form of a scalar field minimally coupled to gravity. 
For the following {\it choice} of the conformal factor
\begin{align}
\Omega^2=F(R)=\frac{\partial f(R)}{\partial R}>0~,
\label{choice_conformal}
\end{align}
the Eq.~\eqref{fR_lagr} becomes
\begin{align}
\mathcal{L}_E = \sqrt{-g_E}\bigg[ \frac{1}{2\kappa^2}R_E - \frac{1}{2}g_E^{\mu\nu}\partial_{\mu}\hat{\phi}\partial_{\nu}\hat{\phi} - V_E(\hat{\phi}) \bigg]~, 
\label{ein_fr}
\end{align}
where, \begin{align}
\hat{\phi} = \frac{1}{\kappa}\sqrt{\frac 3 2 }\ln F.
\label{can1} 
\end{align}
In Eq.~\eqref{ein_fr}, we have dropped a surface term that vanishes at the boundaries. Now the potential function for the field is given by
\begin{align}
V_E(\hat{\phi})=\frac{FR -f(R)}{2\kappa^2 F^2}~.
\label{fr_pot}
\end{align}
Eqs.~\eqref{ein_fr},\eqref{fr_pot} show that any $f(R)$ theory is dynamically equivalent to a minimally coupled scalar field with a potential function determined by the form of $f(R)$. This scalar field is responsible for driving inflation. In the next subsections, we are going to demonstrate that any $f(R)$ theory can be reformulated with the desired pole structure in the kinetic term of the scalar degree of freedom. But whether the theory shows attractor behaviour or not depends on the potential function, that is uniquely determined by the $f(R)$ function. 


\subsection{Relating $f(R)$ Theories to $\xi$-attractor}
Let us now investigate when attractor properties exists for an $f(R)$ theory in its Jordan frame description. For our purpose, we can write the action of Eq.~\eqref{fR_lagr} as \cite{DeFelice:2010aj}
\begin{align}
  S= \frac{1} {2\kappa^2}  \int d^4x\sqrt{-g} \bigg[ F(\phi)(R-\phi) + f(\phi) \bigg] ~.
\label{new_form}
\end{align}
The equation of motion of the scalar field $\phi$ yields $R=\phi$, and it is clear that the above action describes the same theory given by Eq.~\eqref{fR_lagr}. With the identification of $\Omega^2=F(R)$, the action becomes
\begin{align}
  S = \int d^4x\sqrt{-g} \bigg[ \frac{\Omega^2(\phi)R}{2\kappa^2} ~ - ~ \bigg(\frac{F(\phi)\phi - f(\phi)}{2\kappa^2}\bigg)  \bigg]~.
\end{align}
Comparing this theory with what has been defined earlier in Eq.~\eqref{attr_jr}, we find that the resulting structure of the theory is analogous to a $\xi$-attractor with $K_J(\phi)=0$ and $V_J = (F(\phi)\phi - f(\phi))/(2\kappa^2)$ in the Jordan frame. So the behaviour of the resulting potential in this theory is now dependent upon the functional form of $f(R)$. The important difference of this theory with the corresponding $\xi$-attractor is that whereas for the cosmological $\xi$-attractor some particular choices of the Jordan frame potential show attractor like predictions, here the attractor property will rely upon the choice  $f(R)$. 

To have a more clearer picture, we go to the Einstein frame. In this frame, in terms of the variable $\Omega$ (conformal factor) we obtain
\begin{align}
S=\int d^4x \sqrt{-g_E} \bigg[  \frac{1}{2\kappa^2}R_E\! -\! 3\frac{(\partial\Omega)^2}{\Omega^2}\! -\! \frac{V_J(\Omega(R))}{\Omega^4} \bigg]~.
\end{align}    
As we are interested in the nature of the potential at large positive field values, the variable $\Omega^2$ also becomes large. By using the simple transformation $\Omega^2 \rightarrow 1/\rho$, the above Lagrangian transform to Eq.~\eqref{ein_att_approx_rho} with pole at $\rho = 0$. 
Thus in terms of $\Omega^2$, the kinetic term in the Einstein frame has a second order pole. However, now we can not simply choose $V(\Omega)$ so as to make it, for an example, $V_J \propto(\Omega^2-1)^2$. Let us therefore investigate some specific form of $f(R)$ that leads us to attractor like predictions. As we are only interested in understanding the asymptotic behaviour of the potential, here we will not be explicitly carful about dimensionful constants except for one case. The following analysis is complimentary to the discussion in Ref \cite{Broy:2014xwa}  where an investigation has been done to see how functional form $f(R)$ changes when small distortions are made to the case of asymptotic flat potential. 

\subsection*{Case (a): $f(R)\sim R+R^2$} 
This is the famous Starobinsky model \cite{inflation1}. Here we are not careful about the exact coefficients as we we are interested in finding the asymptotic behaviour of the potential. The predictions of this model indeed show attractor nature of Eq.~\eqref{attarctor_observation}. Here we are looking at this model through the non-canonical structure in the Lagrangian density in the Einstein frame. Solving for the Ricci scalar from Eq.~\eqref{choice_conformal} and expressing the potential function in terms of the canonical field we get $R=\frac{\Omega^2 - 1}{2}$, and it gives 
\begin{align}
V(\phi) \simeq (1-\Omega^{-2})^2 \simeq (1-\rho)^2
\label{eq_r2}
\end{align}
In terms of the non-canonical field $\rho$, the Lagrangian has a second order pole in the kinetic term, and the potential is finite positive at the position of the pole $\rho \rightarrow 0$. 
When written in terms of the canonical field as Eq.~\eqref{starobinsky_pot}, the potential asymptotes to the constant value which is equal to the value at the position of the pole. In terms of $\hat{\phi}$ however this pole shifts to infinity and at large $\hat{\phi}$, $V(\hat{\phi})$ approaches to an exponentially flat plateau. The form of the potential remains the same as in the standard case. 

\subsection*{Case (b): $f(R)\sim R+R^3$}
In this case, we get $R=\sqrt{\frac{\Omega^2 - 1}{3}}$ and the potential function
\begin{align}
V(\rho) \simeq\bigg(\frac{1}{\Omega^{2/3}}-\frac{1}{\Omega^{8/3}}\bigg)^{3/2} \simeq\bigg(\rho^{1/3}-\rho^{4/3}\bigg)^{3/2}
\end{align}
We see that both at the position of $\rho = 0(\Omega\to\infty)$, and $\rho = 1(\Omega=1)$, the potential vanishes, and in this case, the potential makes a local maximum that is unsuitable for asymptotically flat potential. In fact, the potential in terms of the canonically normalized field looks like 
\begin{align}
V \simeq e^{-2\sqrt{2/3}\kappa\,\hat{\phi}}(e^{\sqrt{2/3}\kappa\,\hat{\phi}}-1)^{3/2}
\end{align}
For $\rho\to0$ or at large value of the canonical field $\hat\phi$, the potential vanishes because of an overall exponential factor - no suitable vacuum energy to drive inflation. Hence no attractor solution is possible.

\subsection*{Case (c): $f(R) = R+aR^2+ bR^3$}
\begin{figure}[h!]
    \centering
    \includegraphics[width=7.4cm]{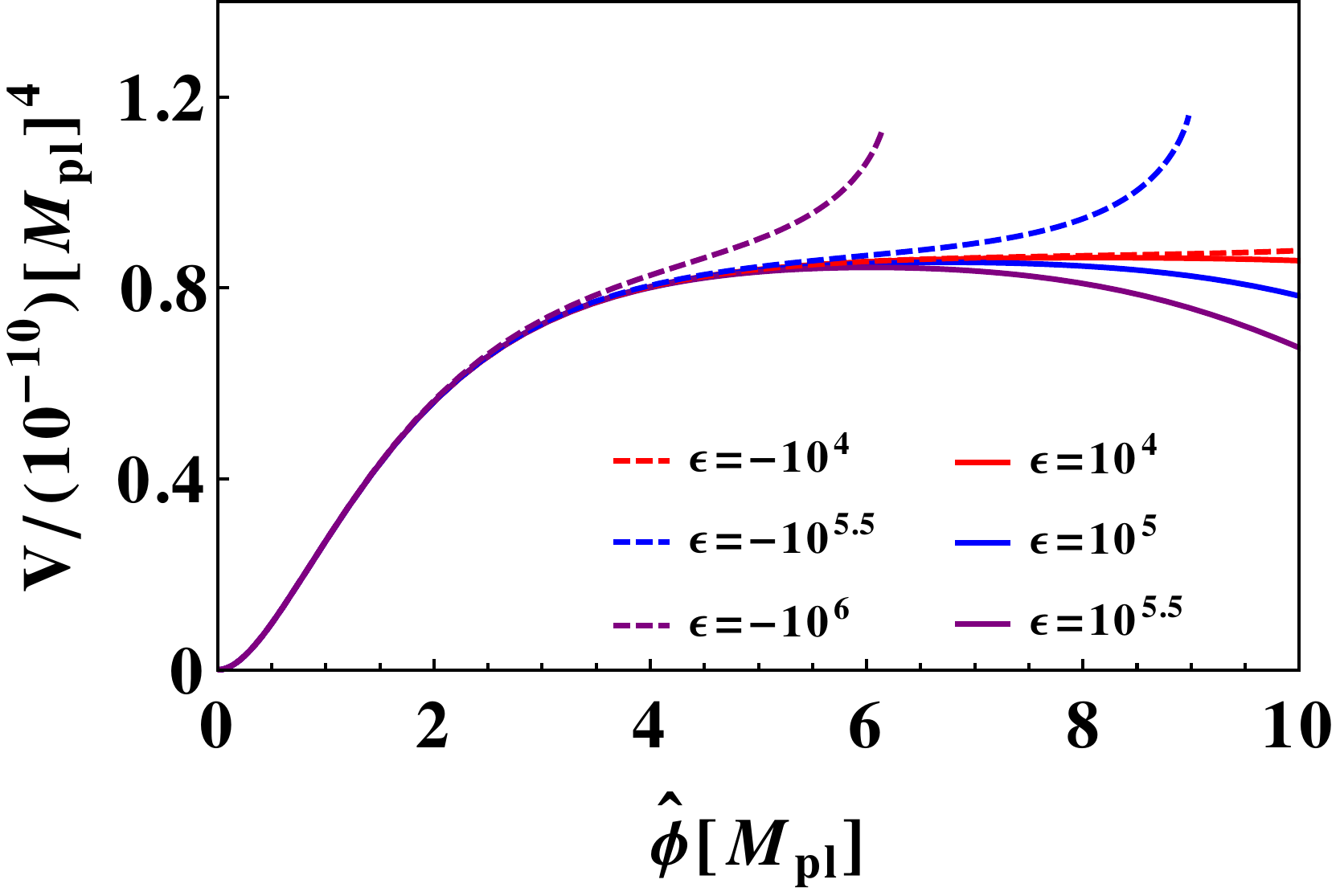} 
    \caption{Einstein frame potential with respect to the canonically normalized field $\hat{\phi}$ for various choices of the parameter ratio $\epsilon=b/a$, taking  $a\sim10^9$}
\label{figr_1}
\end{figure}

Here by solving for $R$ in terms of $\Omega$ we get, 
\begin{align}
R=\frac{-1+\sqrt{1+3\frac{\epsilon}{a}(\Omega^2 -1)}}{3\epsilon}
\end{align}
The potential now depends upon two dimensionful parameters $a$ and $\epsilon$, where $\epsilon=b/a$. We here want to investigate the manner in which the predictions of this model are going to be affected due to presence of the correction term $bR^3$. The Einstein frame potential for this $f(R)$ theory is, 
\begin{align}
V_E=a\frac{(1\!-\!\sqrt{1+3\frac{\epsilon}{a}(\Omega^2\! -\! 1)}\!)^2\,(1\!+\!2\sqrt{1\!+\!\frac{\epsilon}{a}(\Omega^2\! - \!1)}\!)}{54    \epsilon^2 \Omega^4}
\end{align} 
It is easy to check that in the limit $\epsilon\to0$ we get back to the results of Eq.~\eqref{eq_r2} of \textit{Case (a)}. The Fig.~\ref{figr_1} shows how the nature of the potential with respect to the canonical field $\hat{\phi}$ changes as we vary the parameter $\epsilon$ for a given value of the parameter $a$.
The potential depicts different behaviour according to the sign of the $\epsilon$ parameter. We can see that as we keep on increasing $|\epsilon|$ beyond $10^4$ the strength of $R^3$-term begins to dominate. For $\epsilon>0$ the asymptotic behaviour of the potential is such that it gradually looses its height at large field values. This is in contrast to what is found in case of attractor type potentials. However, for $\epsilon<0$ the potential is real only if $\Omega > \sqrt{1-\frac{a}{3\epsilon}}$. Moreover for negative values of $\epsilon$ the potential develops a steep rising branch  \cite{Berkin:1990nu}. As we lower the value of $\epsilon$ the steep branch appears at lower values of $\hat{\phi}$. It turns out that in this case decreasing $\epsilon$ beyond $-10^6$ would not allow  to have 60-efolds of inflation. 

\begin{figure}[h!]
    \centering
    \includegraphics[height=7.4cm]{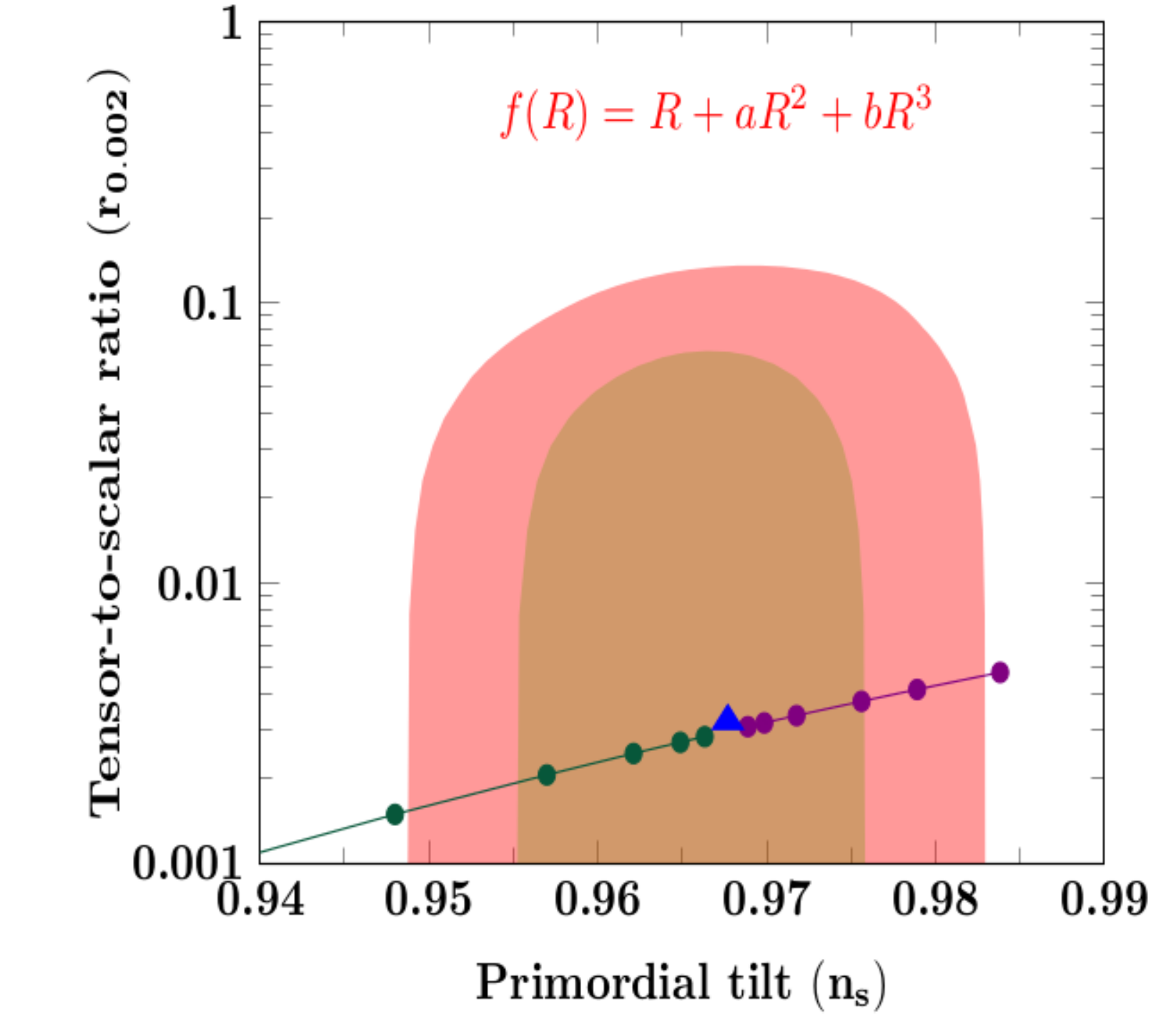} 
             \caption{Plot showing the variations of $n_s$-$r$ with the $68\%$ and $95\%$ CL contours from 2015 Planck data. With decreasing $b$ predictions approaches to Starobinsky model shown here by blue triangle.} 
\label{arbr2}
\end{figure}

Observable predictions of this model is shown in Fig.~\ref{arbr2}. The above figure shows variations of the spectral tilt $n_s$ and tensor-to-scalar ratio $r$ with respect to the parameter $\epsilon$. The plot consists of two branches. In the left branch the green coloured dots indicate points for $\epsilon$ ranging from $10^4$ to about $10^6$ (right to left). The diamond indicates the usual attractor point. It turns out that keeping $a\sim10^9$ (fixed by the amplitude of scalar curvature perturbation) the predictions lie within the PLANCK $2\sigma$ contours as long as $\epsilon\sim10^5$. Beyond that value of $\epsilon$ the strength of $R^3$ term is such that it will violate the spectral index constraint. In the right branch of the plot the violet data points correspond to the range $-10^6<\epsilon<-10^4$. The requirement of real potential restricts $\epsilon$ being larger than $-10^6$, and hence constrains the range of e-folds.

\subsection*{Case (d): $f(R)\sim R+R^n$} 
Here $n$ is a finite integer and $R = \big[\frac{\Omega^2 -1}{n}\big]^{\frac{1}{n-1}}$. Therefore, in terms of the $\Omega$ variable the potential function can be written as,
\begin{align}
V(\Omega) & = \frac{(n-1)\big(\frac{\Omega^2 -1}{n}\big)^{\frac{n}{n-1}}}{\Omega^4} \nonumber \\ 
&\simeq \left( 1 - \rho \right)^{\frac{n}{n-1}}\rho^{\frac{n-2}{n-1}} 
\end{align}  
The above form of the potential function is a generalization of the Starobinsky model when $n=2$, for which it approaches to a constant at at smaller values of $\rho$ (or equivalently at asymptotically large values of $\Omega$). However, for $n>2$ the potential has an overall $\rho$-dependence due to the $\rho^{\frac{n-2}{n-1}}$ factor. Hence at large values of $\hat\phi$ the potential vanishes.

In summary, we can think of the corrections to Einstein gravity in a $f(R)$ theory as equivalent to the higher order corrections in the conformal factor of the theory. For $f(R)$ theory, the conformal factor term is appearing like $\Omega^2(\phi)R = F(R)R =(1+c_0\phi+c_1\phi^2+\ldots)R$, since $\phi=R$ following from Eq.~\eqref{new_form}. Therefore any modification in the $f(R)$ function amounts to a likewise modification in the non-minimal function of Sec.\ref{Sensitivity_of_attractor_models} . Hence the robustness of attractor model investigated in Sec.\ref{Sensitivity_of_attractor_models} can be directly correlated with the various $f(R)$ models envisaged here. Similar analysis in ~\cite{Odintsov_re01} also shows that the modifications to potential function give corrections which are suppressed for power $n > 3$ in $f(R)$.

\subsection{Relating $f(R)$ theories to $\alpha$-attractor}
In the previous subsection we have written down $f(R)$ gravity in terms of conformal factor variable $\Omega^2$, and that clearly spells out the speciality of $R^2$ in terms of its asymptotic nature. Now we will recast the $f(R)$ Lagrangian directly in the form of $\alpha$-attractor given by Eq.~\eqref{alpha_attractor}. It can be easily done if instead of the choice given by Eq.~\eqref{choice_conformal} we {\it choose}
\begin{align}
F=\frac{\sqrt 6 +\kappa\,\phi}{\sqrt 6 - \kappa\,\phi}~.
\label{new_ch}
\end{align}
Then the kinetic term becomes
\begin{align}
\frac{6\kappa^2}{(6-\kappa^2\phi^2)^2}(\partial\phi)^2~.\nonumber
\end{align}
Note that this choice does not affect the expression for the potential function in Eq.~\eqref{fr_pot}. For a given $f(R)$ one solves for $R$ in terms of $\phi$ using Eq.~\eqref{new_ch}. The Lagrangian density for this choice turns out to be,
\begin{align}
\mathcal{L}=\sqrt{- g}\bigg[ \frac{R_E}{2\kappa^2} - \frac{1}{2}\frac{g_E^{\mu\nu}}{\big(1-\frac{\phi^2\kappa^2}{6}\big)^2}\partial_{\mu}\phi\partial_{\nu}\phi - V(\phi) \bigg].
\label{noncan_fr}
\end{align} 
Thus the theory now has a non-canonical field with a pole of order two in the coefficient of its kinetic term. One can also describe the same theory through a canonically normalized scalar field 
\begin{align}
   \hat \phi = \frac{\sqrt 6}{\kappa}\tanh^{-1}{\frac{\phi \kappa}{\sqrt 6}}
\label{can2}
\end{align}
One can easily verify that the two canonical description given by Eq.~\eqref{can1} and Eq.~\eqref{can2} are exactly equivalent. To feature the attractor properties in the $n_s$-$r$ plane we require in addition a smooth potential function at the location of the pole.


\section{Brans-Dicke theory as attractor models}
\label{BD_section}
In this section we will analyze Brans-Dicke models of inflation explicitly. A Brans-Dicke model is an example of $f(R)$-theory. In fact, $f(R)$-theory in metric formalism can be reformulated to the Brans-Dicke theory with Brans-dicke parameter $w = 0$. We will study here the attractor properties of the generalised Brans-Dicke theory defined by the following Lagrangian density,
\begin{align}
\mathcal{L}_J=\frac 1 2 \phi\,R - \frac 1 2 \frac{\omega}{\phi}\,g^{\mu\nu}(\partial_{\mu}\phi)(\partial_{\nu}\phi) - U(\phi)~,\label{BDgeneral}
\end{align}
where $U(\phi)$ is the potential function. Because of the presence of non-minimal coupling term, the description here is in Jordan frame. Now comparing this Lagrangian with the general conformal attractor in Eq.~\eqref{attr_jr},  we see that $\Omega^2(\phi)=\phi$ and $K_J(\phi)=\frac{\omega}{\phi}$. Switching to the Einstein frame we obtain
\begin{align}
\mathcal{L}_E = \sqrt{-\tilde{g}}\bigg[ \frac1 2 \tilde{R} - \frac{1}{2} \frac{(2\omega+3)}{2\phi^2}(\partial\phi)^2 - \frac{U(\phi)}{\phi^2}\bigg]~.
\label{bd_attr}
\end{align}
The above equation has second order pole in the kinetic term. It is clear that with the proper choice of the potential function $U(\phi)$, we can always construct models whose predictions are converged to the attractor point. For the kinetic term in the Einstein frame to be canonical we define,
\begin{align}
\frac{d\hat{\phi}}{d\phi} = \sqrt{\frac{(2\omega+3)}{2\phi^2}}
\label{can_bd}
\end{align} 
Now we want to see for which choice of the potential function $U(\phi)$, the Brans-Dicke theory gives rise to the attractor like predictions. 
From our previous discussion in Section~\ref{attractor-disc}, we know that the choice of the potential must satisfy the three conditions mentioned there to show attractor properties. As the Lagrangian in eq.~\ref{BDgeneral} is exactly equivalent to that of the $\xi$-attractor, choice of potential in the Jordan frame can be either of the two forms: $f\left[ (\phi -1)^{2n}\right] $ or $ f\left[ \left( \frac{\phi -1}{\phi +1}\right) ^{2n}\right] $. In the Brans-Dicke case, the attractor mechanism is highly sensitive to the form of the higher order corrections to the potential. Even if we add these higher order corrections, the corrections have to take the forms as $a_me^{-c\hat{\phi}}$ to preserve the attractor behaviour of the Lagrangian. See ~\cite{Kallosh:2013tua} for more discussion about the forms of the correction terms in the potential.

We make two following choices: 
\subsection{$\mathbf{U(\phi)=U_0(\phi-1)^2}$}
\noindent where $U_0$ is a potential parameter that is to be fixed from the value of the scalar power spectrum. Even though there is a second order pole at $\phi=0$ in the kinetic term, however in terms of the canonical field obtained from Eq.~\eqref{can_bd}, this pole shifts to infinity. In this case, the potential in terms of the canonical field in the Einstein frame is given by
\begin{equation}
V_E(\hat{\phi}) = U_0(1-e^{-c\hat{\phi}})^2,
\end{equation}
where $c=\sqrt{\frac{2}{2\omega+3}}$.

\begin{figure}[h!]
    \centering
    \includegraphics[height=6.6cm]{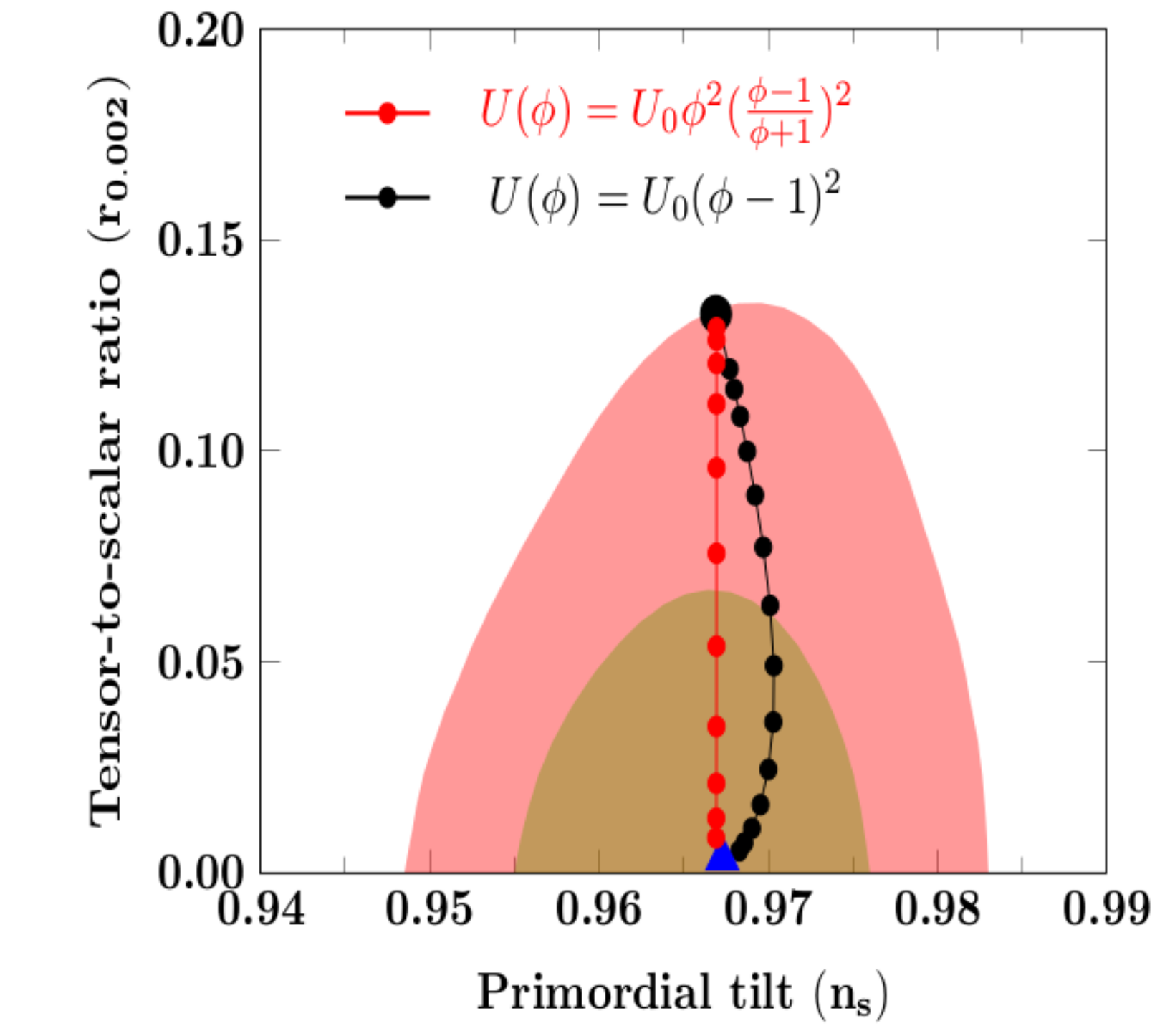} 
             \caption{Plot showing the variatons of $n_s$-$r$ overlated with the $68\%$ and $95\%$ CL contours from 2015 Planck data. With decreasing $\omega$ predictions asymptotically approaches to Starobinsky model shown here by the blue triangle.} 
\label{nsr_bd1}
\end{figure}

Let us now investigate the predictions of this potential in the light of PLANCK 2015 data. The observable predictions
for this potential are shown in Fig~\ref{nsr_bd1}. 
The black dots indicate variations in the inflationary predictions with respect to the Brans-Dicke parameter $\omega$ ranging from $1$ to about $10^6$. The plot depicts that with decreasing the value of $\omega$ (top to bottom) the predictions of this model interpolates between quadratic chaotic inflation and Starobinsky model. Taking amplitude of scalar curvature perturabtions, $A_s$ to lie within the $99.7\%$ CL of the Planck data we fix $U_0\sim10^{-10}$. We now turn to a different choice of the potential function:
\subsection{$\mathbf{U(\phi)=U_0\phi^2\big(\frac{\phi-1}{\phi+1}\big)^2}$}    
\noindent In this case the potential function in the Einstein frame in terms of the canonically normalized inflaton field  is given by 
\begin{align}
V_E(\hat{\phi}) = U_0 \tanh^2 {\frac{\hat{\phi}}{\sqrt{4\omega+6}}}  
\end{align}
This potential is nothing but the simplest generalization of the T-model of $\alpha$-attractor with the identification $\alpha=\frac2 3 \omega +1$ \cite{T-model}. Therefore, in the leading order approximation in the inverse efolds the inflationary predictions are
\begin{align}
n_s = 1-\frac 2 N, \qquad r=\frac{12(1+\frac{2\omega}{3})}{N^2}~.
\end{align}
We numerically solve the dynamics in the above potential, and the observable predictions are plotted in Fig~\ref{nsr_bd1}.



\section{Conclusions and Discussions}
PLANCK 2015 data prefers inflation models with plateau like potential with asymptotic flatness \cite{planck2015, Ijjas:2013vea}. Among many models, the modified gravity model proposed by Starobinsky has attracted a lot of attention due to its observational predictions that are nearly in the middle of 2-$\sigma$ contours of spectral index and tensor-to-scalar ratio plane. A class of cosmological models has been found subsequently whose observational predictions in the $n_s$-$r$ plane are attracted to this Starobinsky value when a parameter of the model is changed continuously. These models, termed as attractor models, draw their attractor properties from certain pole structure of the kinetic term. 

In this work, we have analysed the scalar-tensor theories of gravity in the light of the attractor models. In particular, we work with $f(R)$ gravity models, and recast the models in the form of attractor models. Any particular choice of $f(R)$ automatically fixes the form of the scalar potential function. Therefore, even though any $f(R)$ model can be recasted with the desired form of the kinetic energy with a certain pole structure, only for a certain case it satisfies the required condition for the scalar potential. This behaviour singles out $R^2$ gravity models from any other modifications. Any higher order term does not satisfies the desired asymptotic properties of the potential. We  have analysed inflationary phenomenology when higher order terms in the action is also present. We also look at the Brans-Dicke theory of inflation, and find suitable potential functions that automatically provides the attractor predictions when the Brans-Dicke parameter $w$ is varied appropriately. 

We also analyze the stability of attractor mechanism. Only for a certain choice of the potential function in the Jordan frame, the potential is asymptotically flat with a constant vacuum energy. Any higher order term in the conformal function makes the potential asymptotically zero. But, if the attractor parameter $\xi$ is increased sufficiently, the field range where observable inflation happens remains sufficiently flat, and the predictions return to the Starobinsky attractor point. This shows the robustness of the attractor mechanism. We also discuss how the predictions change when higher order poles are simultaneously present in the kinetic function. In this case, the existence of higher order pole always makes the predictions away from the usual attractor curve for second order pole. If we decrease the residue of the second order pole sufficiently, the effect of the second order pole term becomes subdominant as $\phi_{60}$ remains almost constant. 

The analysis can be extended to different generalised versions of scalar-tensor theories. In particular, it would be interesting to find attractor type solutions for theories with derivative couplings \cite{Amendola:1993uh}, and non-local modifications of gravity \cite{Deser:2007jk}. In these cases, the crucial point is to find suitable conformal transformation that can recast the kinetic energy term with a certain pole structure in the Einstein frame. Once that is achieved, the potential function can be designed to get the attractor solutions. We hope to come back to this issue in future. 
\vspace{5mm}
\subsection*{Acknowledgements}
Sukannya is supported by a Ph.D fellowship from CSIR, Govt. of India. Kumar is supported by a Ph.D fellowship from the DAE, Govt. of India. Koushik is partially supported by a Ramanujan Fellowship from DST (SERB), Govt of India.


\end{document}